\documentclass[twocolumn,showpacs,preprintnumbers,amsmath,amssymb,superscriptaddress,prr,longbibliography]{revtex4-2}
\usepackage{graphicx,here}% Include figure files
\usepackage{dcolumn}% Align table columns on decimal point
\usepackage{bm}% bold math
\usepackage{enumerate}

\newcommand{\myvec}[1]{\mbox{\boldmath $#1$}}

\begin{document}

\preprint{APS/123-QED}

\title{Stroboscopic Time-of-Flight Neutron Diffraction in Long Pulsed Magnetic Fields}

\author{Taro Nakajima}
\email{taro.nakajima@issp.u-tokyo.ac.jp}
\affiliation{The Institute for Solid State Physics, the University of Tokyo, Kashiwa, Chiba, 277-8581, Japan.}
\affiliation{RIKEN Center for Emergent Matter Science (CEMS), Saitama 351-0198, Japan.}
\author{Masao Watanabe}
\affiliation{J-PARC Center, Japan Atomic Energy Agency, Tokai, Ibaraki 319-1195, Japan.}
\author{Yasuhiro Inamura}
\affiliation{The Institute for Solid State Physics, the University of Tokyo, Kashiwa, Chiba, 277-8581, Japan.}
\affiliation{J-PARC Center, Japan Atomic Energy Agency, Tokai, Ibaraki 319-1195, Japan.}
\author{Kazuki Matsui}
\author{Tomoki Kanda}
\author{Tetsuya Nomoto}
\affiliation{The Institute for Solid State Physics, the University of Tokyo, Kashiwa, Chiba, 277-8581, Japan.}
\author{Kazuki Ohishi}
\author{Yukihiko Kawamura}
\affiliation{Neutron Science and Technology Center, Comprehensive Research Organization for Science and Society (CROSS), Tokai, Ibaraki 319-1106, Japan.}
\author{Hiraku Saito}
\affiliation{The Institute for Solid State Physics, the University of Tokyo, Kashiwa, Chiba, 277-8581, Japan.}
\author{Hiromu Tamatsukuri}
\affiliation{J-PARC Center, Japan Atomic Energy Agency, Tokai, Ibaraki 319-1195, Japan.}
\author{Noriki Terada}
\affiliation{National Institute for Materials Science, Sengen 1-2-1, Tsukuba, Ibaraki 305-0047, Japan.}
\author{Yoshimitsu Kohama}
\affiliation{The Institute for Solid State Physics, the University of Tokyo, Kashiwa, Chiba, 277-8581, Japan.}

\begin{abstract}
We present proof-of-principle experiments of stroboscopic time-of-flight (TOF) neutron diffraction in long pulsed magnetic fields. %
By utilizing electric double-layer capacitors, we developed a long pulsed magnet for neutron diffraction measurements, which generates pulsed magnetic fields with the full widths at the half maximum of more than $10^2$ ms. %
The field variation is slow enough to be approximated as a steady field within the time scale of a polychromatic neutron pulse passing through a sample placed in a distance of the order of $10^1$ m from the neutron source. %
This enables us to efficiently explore the reciprocal space using a wide range of neutron wavelength in high magnetic fields. %
We applied this technique to investigate field-induced magnetic phases in the triangular lattice antiferromagnets CuFe$_{1-x}$Ga$_x$O$_2$ ($x=0, 0.035$).

\end{abstract}

\pacs{}% PACS, the Physics and Astronomy
                             % Classification Scheme.
%\keywords{Suggested keywords}%Use showkeys class option if keyword
                              %display desired
\maketitle

\section{INTRODUCTION}

Exploring new quantum states of matter in extreme conditions, such as low temperatures, high magnetic fields, and high pressures, is one of the central topics in condensed matter physics. %
Among them, novel field-induced phases have been extensively investigated; for example, spin-lattice-coupled magnetization plateaus in frustrated spin systems \cite{HalfMagPlateauCdCr2O4_PRL,PRL2004_Penc,NPhys_Matsuda2007,SrCu2(BO3)2_100T}, %
field-induced flop of the spin-driven electric polarization in multiferroics \cite{Kimura_nature,PRL_Ni3V2O8}, %
spin-nematic states in quantum spin systems \cite{Kohama_PNAS_nematic,Nematic_LiCuVO4} etc. %
Neutron scattering is one of the most powerful techniques to study these exotic phenomena because it can probe Fourier-transformed time-space correlation functions of atomic positions and magnetic moments; specifically crystal and magnetic structures are determined by the elastic scattering, and phonons and magnons are measured by the inelastic scatterings. %
However, the highest magnetic field for neutron scattering instruments is limited to approximately 15 T even for state-of-the-art superconducting magnets. %
Although there was a superconducting and non-superconducting hybrid magnet for neutron scattering with the highest magnetic field of 26 T in the Helmholtz Zentrum Berlin (HZB)\cite{HFM_EXED}, it is no longer available since the research reactor in HZB was shut down in 2019. %

An alternative is a pulsed magnetic field\cite{Pulse_Nojiri,Pulse_Nojiri_MnWO4,Pulse_LiNiPO4,ILL_pulseMag}. %
For instance, Nojiri and co-workers successfully observed a magnetic Bragg peak in a field-induced phase of multiferroic MnWO$_4$ by time-of-flight (TOF) pulsed neutron diffraction measurements with pulsed magnetic fields up to 30 T\cite{Pulse_Nojiri_MnWO4}. %
However, the widths of the pulsed magnetic fields are often limited to several milliseconds, which are shorter than a time spread of a polychromatic neutron pulse after flying a typical source-to-sample distance in existing TOF neutron diffractometers, as we explain later. %
In this case, the high magnetic field is achieved only for a limited wavelength range of the neutrons, and this situation makes explorations of new magnetic peaks in the reciprocal space difficult. %

To overcome this problem, in the present study, we have developed a long pulsed magnet for neutron scattering experiments. %
By utilizing electric double-layer capacitors (EDLC)\cite{LongPulse_EDLC}, we successfully generated pulsed magnetic fields whose full widths at the half maximum exceed 100 ms. %
We established the low-temperature neutron scattering environment with this long pulsed magnet, and performed stroboscopic TOF neutron diffraction measurements on the triangular lattice antiferromagnets CuFe$_{1-x}$Ga$_x$O$_2$ ($x=0, 0.035$). %
By virtue of the extended time scale of the magnetic field, we can use a wide wavelength range of the incident neutron beam, which enables us to map out the neutron diffraction intensities in the reciprocal space in high magnetic fields. %

\begin{figure}[t]
\begin{center}
\includegraphics[clip,keepaspectratio,width=8.0cm]{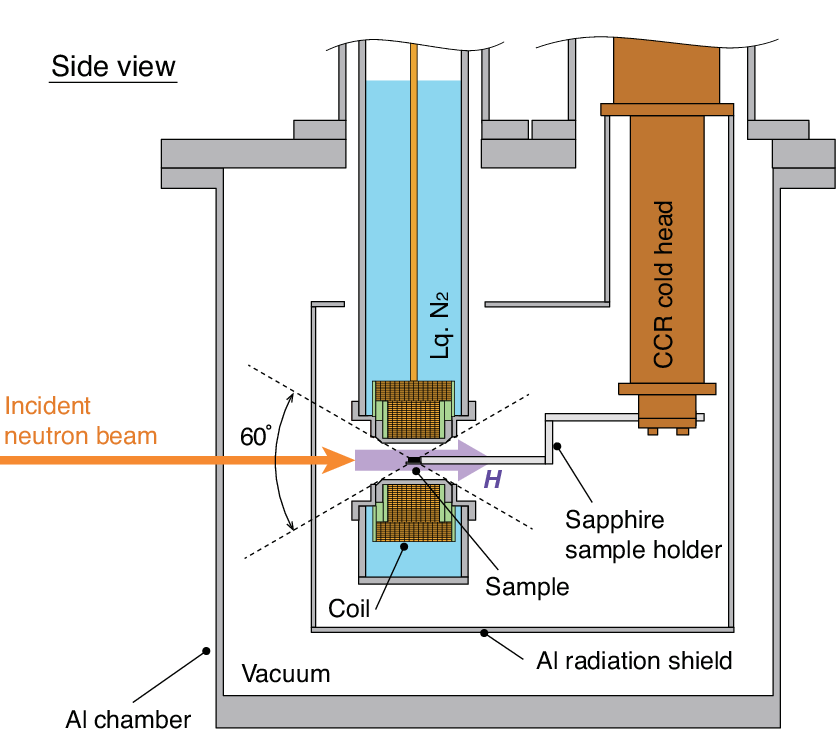}
\caption{Schematic illustration of the side view of the long pulsed magnet and CCR in the Al vacuum chamber. }
\label{cryo}
\end{center}
\end{figure}

CuFeO$_2$ has been extensively investigated as a geometrically frustrated system \cite{Mitsuda_1991,Mekata_1993}. %
Crystal structure of this compound belongs to the space group of $R\bar{3}m$, and consists of triangular lattice layers of magnetic Fe$^{3+}$ ions, which are rhombohedrally stacked along the $c$ axis. %
In zero magnetic field, this compound exhibits a collinear four-sublattice (4SL) antiferromagnetic order, in which the spins are arranged to form an up-up-down-down sequence running along the [110] direction, below 11 K. %
The spins in the adjacent layers are antiferromagnetically aligned, and thus the magnetic modulation wave vector ($q$-vector) is described as $q=(\frac{1}{4},\frac{1}{4},\frac{3}{2})$. %
The 4SL antiferromagnetic order breaks the threefold rotational symmetry about the $c$ axis of the crystal, and thus results in three magnetic domains corresponding to the $q$-vectors of $q=(\frac{1}{4},\frac{1}{4},\frac{3}{2})$, $(-\frac{1}{2},\frac{1}{4},\frac{3}{2})$ and $(\frac{1}{4},-\frac{1}{2},\frac{3}{2})$, which are interconverted to each other by the threefold rotation operation. %
Previous x-ray diffraction studies revealed that each magnetic domain exhibits a monoclinic lattice distortion associated with the $q$-vector\cite{Terada_CuFeO2_Xray,Ye_CuFeO2}. %
For convenience, we employ the hexagonal basis when referring $q$-vectors and positions in the reciprocal space in the rest of this paper, although they could be properly indexed using the monoclinic basis. %

By applying a magnetic field along the $c$ axis at low temperatures, CuFeO$_2$ exhibits field-induced successive phase transitions\cite{Mitsuda_2000_H-T,Mitsuda_2000,Petrenko_2000}. %
The first-field-induced phase is known to have an incommensurate $q$-vector and spin-driven ferroelectricity\cite{Kimura_CuFeO2}. %
The second field-induced phase exhibits a $\frac{1}{5}$-magnetization plateau. %
In this paper, we refer to these phases as ferroelectric incommensurate-magnetic (FE-ICM) and $\frac{1}{5}$-plateau phases, respectively. %
Both phases were studied by previous neutron scattering experiments with vertical-field superconducting magnets\cite{Mitsuda_2000,Petrenko_2000}. %
However, the accessible scattering plane was limited to the $(h,k,0)$ plane, where the fundamental magnetic reflections were not observed due to the finite $l$ component of the $q$-vectors. %
It is also known that the critical field between the 4SL and FE-ICM phases can be reduced to zero by substituting a small amount of nonmagnetic ions for Fe$^{3+}$ \cite{Seki_PRB_2007,Kanetsuki_JPCM}. %
Thus, the magnetic structure of the FE-ICM phase was investigated by neutron diffraction measurements on CuFe$_{1-x}A_x$O$_2$ ($A=$Al, Ga), and determined to be the incommensurate screw-type magnetic structure \cite{SpinNoncollinearlity,CompHelicity}. % 
Although the nonmagnetic-substitution-induced phase was considered to be the same as the field-induced FE-ICM phase in pure CuFeO$_2$, the direct evidence is still lacking because the fundamental magnetic Bragg reflections in the field-induced phase of pure CuFeO$_2$ were not directly observed. %

Field-induced magnetic phase transitions in CuFe$_{1-x}$Ga$_x$O$_2$ were also studied by means of magnetization, electric polarization, ESR and heat capacity measurements\cite{Seki_CFGO,CFGO_highfield_ESR,CFGO_heatcapacity}. %
At $x=0.035$, where the ground state is already replaced with the FE-ICM phase, the first field-induced phase is considered to be the $\frac{1}{5}$-plateau phase. %
However, the plateau in magnetization is smeared with increasing $x$, and therefore the direct observation of the $q$-vectors by neutron diffraction is necessary to unambiguously determine the spin arrangements in the field-induced phase of the $x=0.035$ sample. %

\begin{figure}[t]
\begin{center}
\includegraphics[clip,keepaspectratio,width=8.5cm]{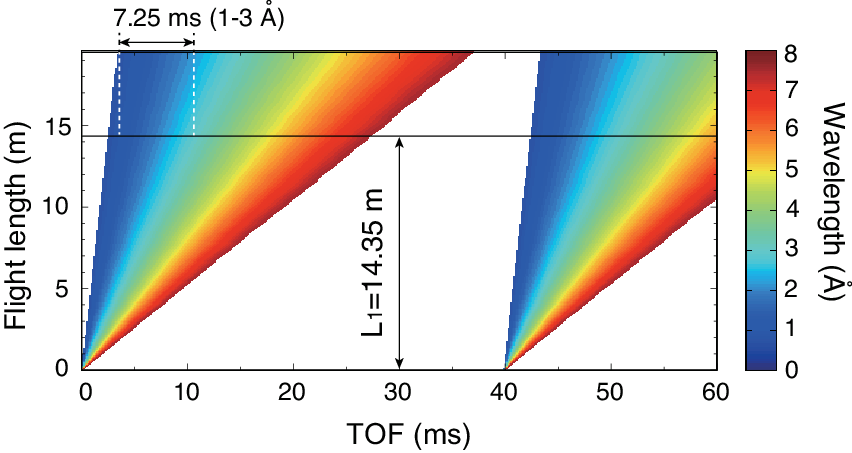}
\caption{The TOF diagram of TAIKAN in MLF. The color map shows the wavelengths of the incident neutrons.  }
\label{TOF}
\end{center}
\end{figure}

\section{EXPERIMENTAL DETAILS}

\subsection{Sample, Cryostat and Magnet}
A single crystal of CuFeO$_2$ was grown by the floating zone method \cite{Zhao_FZ}, and cut into a cylinder shape with the diameter and length of 5 and 6 mm, respectively. %
The sample was glued on a sapphire sample holder, which was connected to the cold head of a $^4$He closed-cycle refrigerator (CCR). %

We prepared a pulsed magnet consists of inner solenoid of a 18-layer and 11 turn and outer solenoid of 10-layer and 16 turn; both were wound with a copper wire having a rectangular cross-section of $2.4\times 1.0$ mm$^2$. %
The magnet produces a horizontal magnetic field, and has two conical windows with an opening angle of 60$^{\circ}$, as shown in Fig. \ref{cryo}. %
The inner bore of the magnet is 20 mm in diameter. %
The magnet was cooled by liquid N$_2$, and was thermally isolated from the sample by vacuum. %
The coil inductance, coil resistance, and the field factor at 77 K are 3.9 mH, 57.5 m$\Omega$, and 7.85 T/kA, respectively.
The sample was set at the field center. %
The direction of the magnetic field was parallel to the $c$ axis, namely perpendicular to the triangular lattice plane. %
The [110] direction was selected to be the horizontal direction perpendicular to the field direction. %
The magnet and sample were put in an Al-chamber, which was also pumped to high vacuum. %
Note that the Al-chamber and CCR were originally designed for the short pulsed magnet system in the Materials and Life-science experimental Facility (MLF) of J-PARC \cite{Watanabe_ShortPulse}, and are also compatible with the long pulsed magnet. % 

A single crystal of CuFe$_{1-x}$Ga$_x$O$_2$ ($x=0.035$) was also grown by the floating zone method \cite{Zhao_FZ}, and cut into a rectangular shape with the dimensions of 4.9$\times$2.9$\times$3.5 mm$^3$. %
The sample environments and the alignment of the crystal were the same as those for the $x=0$ sample. 

\subsection{Time Structure of A Neutron Pulse}
The neutron diffraction experiments were carried out at the small-angle and wide-angle neutron TOF diffractometer TAIKAN in MLF of J-PARC \cite{J-PARC_TAIKAN}. %
Figure \ref{TOF} shows a typical TOF diagram for TAIKAN. %
Since neutron velocity is inversely proportional to its wavelength, a sharp polychromatic neutron pulse created at the neutron source has a wavelength spread after flying the source-to-sample distance ($L_1$), which is 14.35 m at TAIKAN. %
Taking into account that the interval between two adjacent neutron pulses is 40 ms in MLF, TAIKAN normally uses the wavelength range from 0.7 to 7.7 \AA. %
In the present study, we focus on the wavelength range from 1 to 3 \AA\  to measure magnetic Bragg reflections of CuFe$_{1-x}$Ga$_x$O$_2$, as we show in the following sections. %
This wavelength range corresponds to the time spread of 7.25 ms at the sample position, as shown in Fig. \ref{TOF}. %
If the width of the pulsed magnetic field is much larger than this time spread, we can fully utilize the wavelength range to  map out the neutron diffraction intensities in the reciprocal space. %

\begin{figure}[t]
\begin{center}
\includegraphics[clip,keepaspectratio,width=8.5cm]{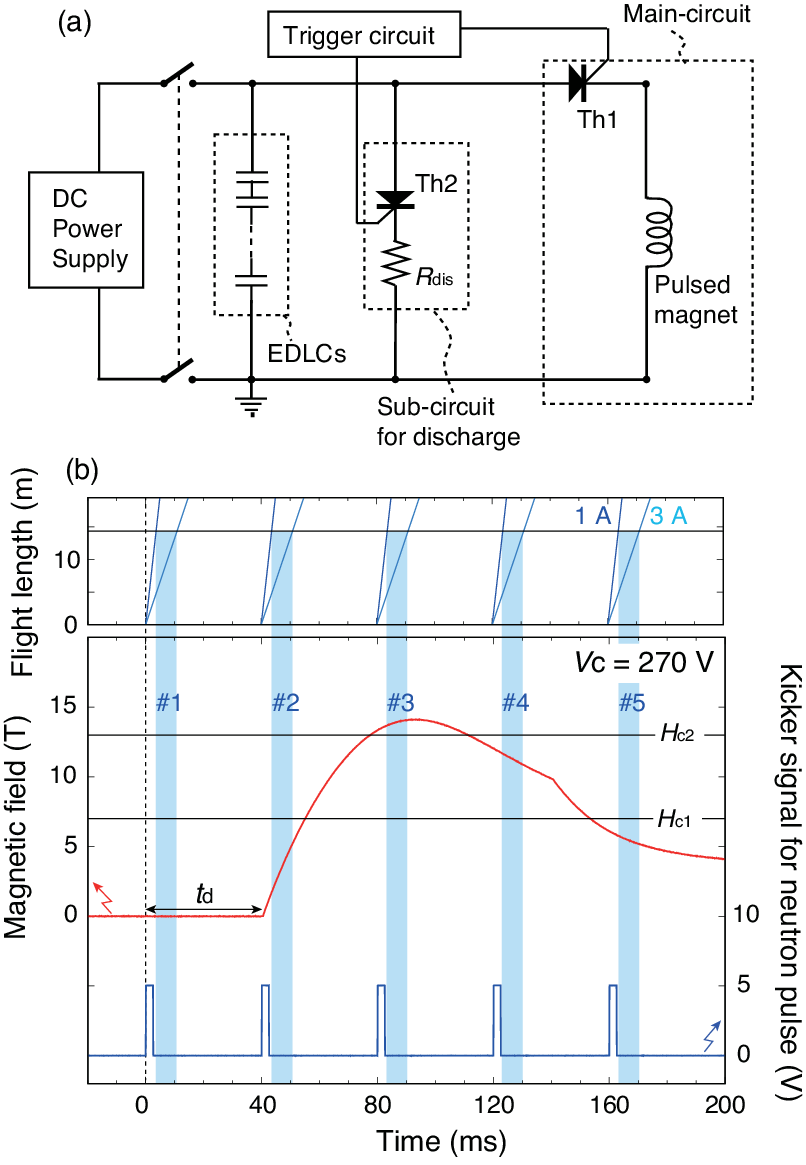}
\caption{(a) Circuit diagram of the EDLC-based pulse power supply for the generation of long pulsed magnetic fields. %
(b) Comparisons among the TOF diagram, the temporal profile of the long pulsed magnetic field generated with a charging voltage of 270 V, and the kicker signals for neutron pulse generation. %
The neutron pulses are triggered by the rising edges of the kicker signals. }
\label{MagPulse}
\end{center}
\end{figure}

\begin{figure}[t]
\begin{center}
\includegraphics[clip,keepaspectratio,width=8.0cm]{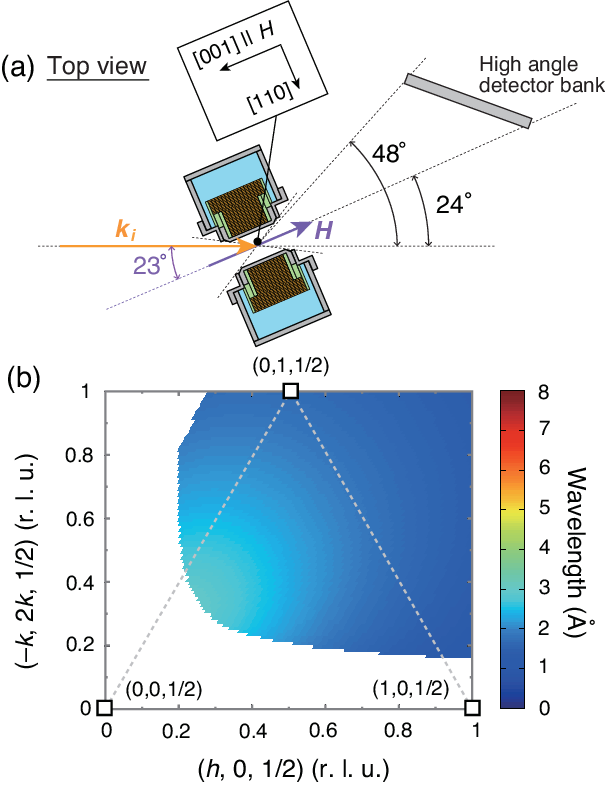}
\caption{(a) A schematic illustration of the top view of the instrument layout in the present experiment. %
(b) A color map showing the neutron wavelengths for measuring intensities on the $(h,k,\frac{1}{2})$ plane in the scattering geometry shown in (a).  }
\label{Geometry}
\end{center}
\end{figure}

\subsection{Generation of Long Pulsed Magnetic Fields}

We have constructed an electrical circuit for generating long pulsed magnetic fields which consists of an EDLC bank, a DC power supply, a trigger circuit, two thyristors (Th1 and Th2), a discharge resistance ($R_{\rm dis}$), and a pulsed magnet, as shown in Fig. \ref{MagPulse}(a). %
The EDLC bank consists of a series connection of 120 EDLC cells (Nippon Chemi-Con) and has a total capacitance of 30 F and an internal resistance of $37$ m$\Omega$ with a maximum charged voltage of 300 V. %
The DC power supply (TDK-Lambda Americas) was used to charge the EDLC bank. %
The trigger circuit controls the timing of closing Th1 and Th2. %
The current flow through the pulsed magnet, which is proportional to the field strength, starts at the timing of closing Th1. 

In Fig. \ref{MagPulse}(b), we show a temporal profile of the long pulsed magnetic field generated by a charging voltage of $V_c=270$ V. %
The width of the field pulse is more than one order of magnitude larger than the time spread corresponding to the wavelength range from 1 to 3 \AA\, which is indicated by the light-blue rectangles in Fig. \ref{MagPulse}(b). %
This means that the long pulsed magnetic field can be approximated as a steady field within the time scale of the neutron pulses. %
We note here that there is a sudden drop of the magnetic field upon the field decreasing process. %
At this point, Th2 is intentionally closed so that the electric current flows not only to the main-circuit but also to the sub-circuit for discharge. % 
While the intentional reduction of the field strength slightly reduce the width of the pulsed field, this suppresses unnecessary heating of the coil, and thus reduces a waiting time for applying next magnetic field pulse, which was approximately 9 minutes at maximum in the present experiment. %

\begin{figure*}[t]
\begin{center}
\includegraphics[clip,keepaspectratio,width=18.0cm]{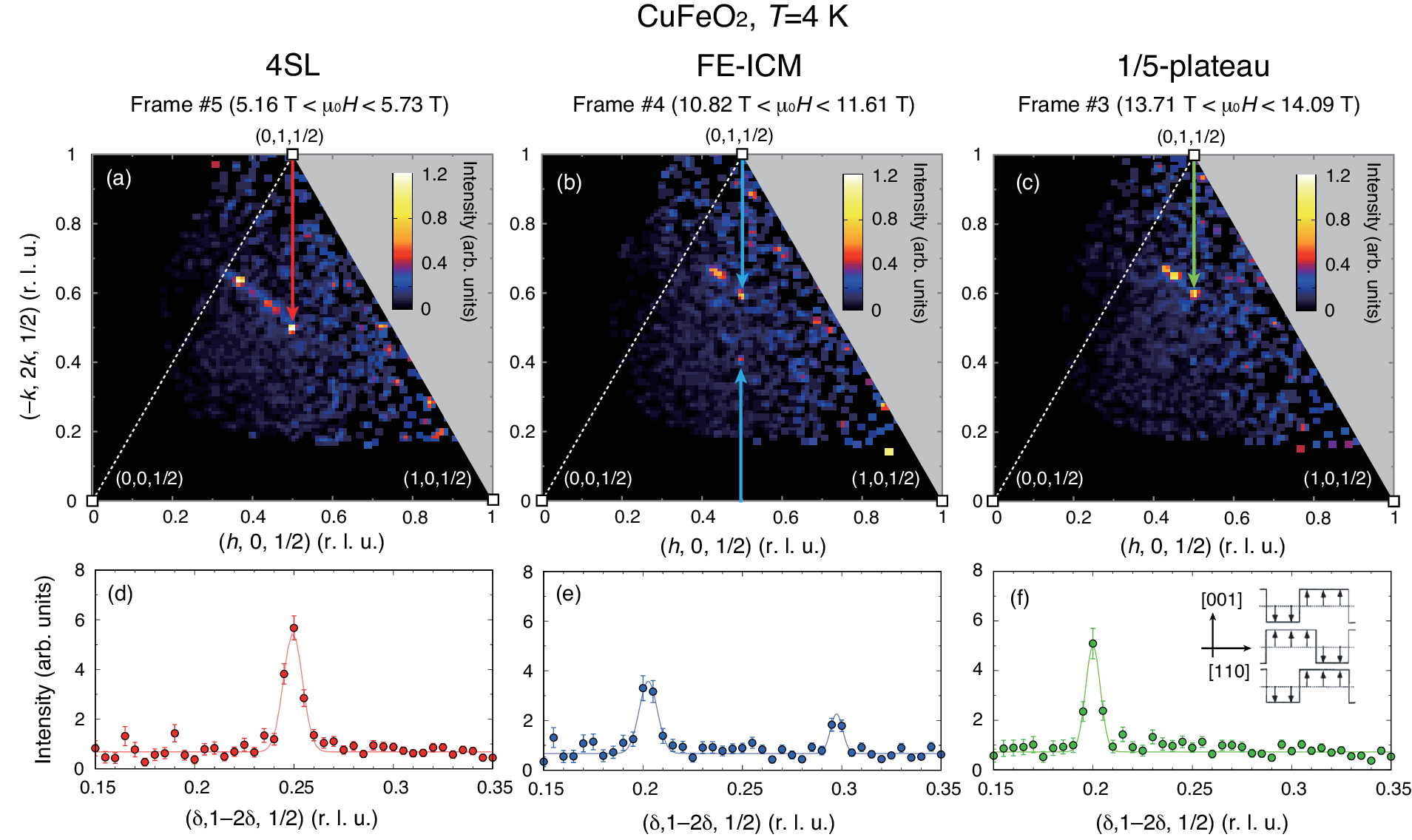}
\caption{Stroboscopic neutron diffraction results on CuFeO$_2$. [(a)-(c)] Intensity maps on the $(h,k,\frac{1}{2})$ plane in the (a) 4SL, (b) FE-ICM, and (c) $\frac{1}{5}$-plateau phases. %
These data are obtained from the \#5, \#4, and \#3 frames, respectively, measured with the long pulsed magnetic field generated with the charging voltage of 270 V, as shown in Fig. \ref{MagPulse}. %
The application of the magnetic field was repeated for 122 times. %
Solid arrows indicate the positions of the magnetic Bragg peaks in each phase. %
[(d)-(f)] Line profiles along the $(\delta,1-2\delta,\frac{1}{2})$ direction in the (d) 4SL, (e) FE-ICM, and (f) $\frac{1}{5}$-plateau phases. %
Inset of (f) shows the spin arrangement of the $\frac{1}{5}$-plateau phase cited from Ref. \onlinecite{Mitsuda_2000}. }
\label{IntMap}
\end{center}
\end{figure*}

\begin{figure*}[t]
\begin{center}
\includegraphics[clip,keepaspectratio,width=18.0cm]{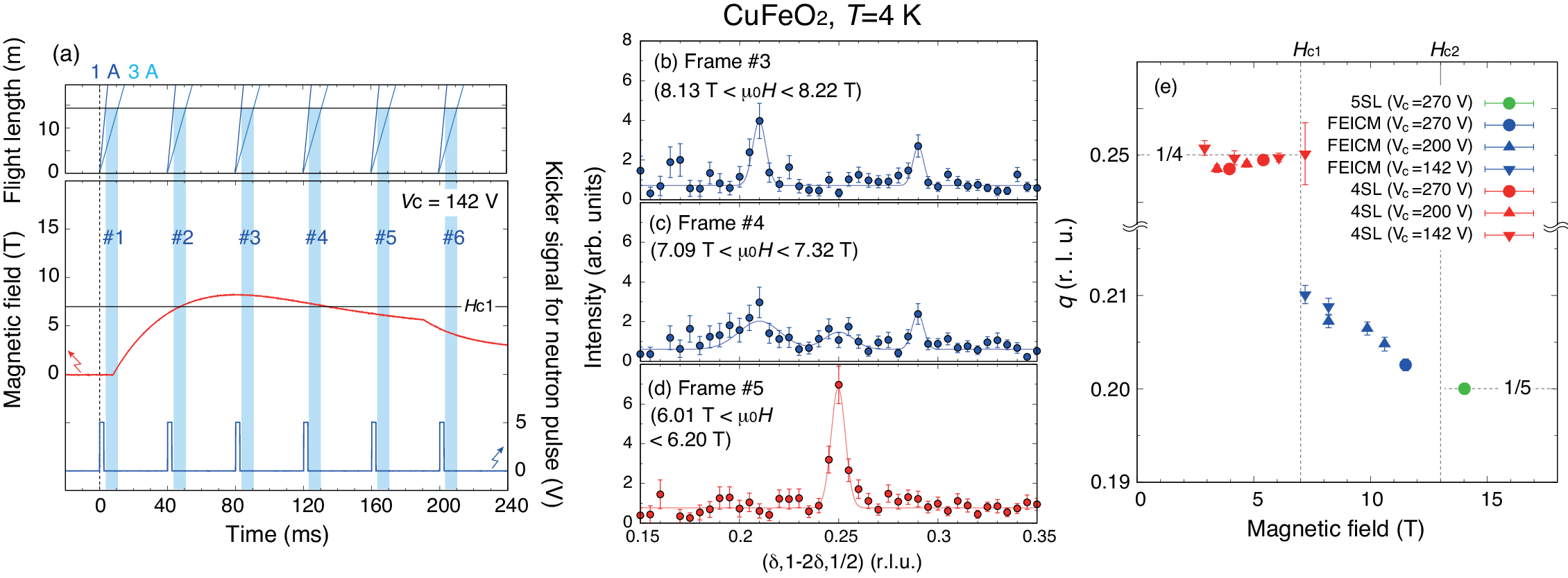}
\caption{(a) Comparisons among the TOF diagram, the temporal profile of the long pulsed magnetic field generated with a charging voltage of 142 V, and the kicker signals for neutron pulse generation. %
[(b)-(d)] Line profiles along the $(\delta,1-2\delta,\frac{1}{2})$ direction in CuFeO$_2$ measured by the frame (b) \#3, (c) \#4, and (d) \#5, shown in (a). %
(e) Field dependence of the magnetic modulation wave number $q$ in CuFeO$_2$, which was determined from the peak position of the line profile along  $(\delta,1-2\delta,\frac{1}{2})$.  }
\label{Hdep}
\end{center}
\end{figure*}

\subsection{Stroboscopic TOF Neutron Diffraction}
Despite the extended time scale of the pulsed magnetic fields, the intensity of one neutron pulse is still not strong enough to precisely measure the reflections from the sample. %
We thus repeated the application of a magnetic field pulse to accumulate the data. %
The application of a field pulse was triggered by one of the kicker signals for the neutron pulse generation. %
The data measured by each neutron pulse were numbered starting from this trigger, as shown in Fig. \ref{MagPulse}. %
In the following, we refer to these numbers as frame numbers. %
We also introduced a delay time $t_d$ to adjust the peak position of the magnetic field with respect to the neutron pulses. %
After repeating the field applications, we extracted and accumulated the data having the same frame number by UTSUSEMI software \cite{UTSUSEMI}. %
Finally, we obtained stroboscopic neutron diffraction patterns in the long pulsed magnetic fields. %

Figure \ref{Geometry}(a) shows the experimental configuration. %
The angle between the incident neutron beam and the magnetic field was set to be 23 deg. %
We mainly used the high-angle detector bank of TAIKAN, which has a horizontal coverage of the scattering angle from 24$^{\circ}$ to 48$^{\circ}$. %
Note that we can also measure reflections out of the horizontal plane owing to the vertical coverage of the high-angle detector bank. %
Figure \ref{Geometry}(b) is the color map of the neutron wavelengths for measuring intensities on the $(h,k,\frac{1}{2})$ plane in the configuration mentioned above. %
Specifically, the scattering vector $\myvec{Q}(=\myvec{k}_i-\myvec{k}_f)$, where $\myvec{k}_i$ and $\myvec{k}_f$ are wavevectors of the incident and scattered neutrons, respectively, for an elastic scattering event is determined by the wavelength and the direction of the scattered neutron. %
When the scattering vector is in the colored region in Fig. \ref{Geometry}(b), the elastic scattering condition is satisfied with the wavelength indicated by the color, and the direction of the scattered neutron is covered by the high-angle detector bank. %
In this region, we can observe satellite reflections with $q$-vectors indexed as $(q,-2q,\frac{3}{2})$, which is equivalent to $(q,q,\frac{3}{2})$ under the threefold rotation about the $c$ axis, from the reciprocal lattice point of $(0,1,-1)$. %  
The wavelength range to cover this region is from 1 to 3 \AA.

\section{RESULTS AND DISCUSSIONS}

\subsection{CuFeO$_2$}
We performed the stroboscopic neutron diffraction in long pulsed magnetic fields on CuFeO$_2$ at 4 K, at which the lower critical fields of the FE-ICM and 1/5-plateau phases, namely $H_{c1}$ and $H_{c2}$, are 7 and 13 T, respectively. %
We repeated the application of the long pulsed magnetic field with the charging voltage of 270 V for 122 times, and then extracted the diffraction patterns of \#3, \#4 and \#5 frames, which correspond to the 1/5-plateau, FE-ICM, and 4SL phases, respectively, as shown in Fig. \ref{MagPulse}. %
The intensities are normalized to the incident neutron flux taking into account its wavelength dependence, and mapped onto the $(h,k,\frac{1}{2})$ plane after integrating them with respect to $l$. The integration range for $l$ is from $0.45$ to $0.55$. %

Figure \ref{IntMap}(a) shows the intensity map of the 4SL phase. %
We observed a magnetic Bragg reflection at $(\frac{1}{4},\frac{1}{2},\frac{1}{2})$, which is assigned as a satellite reflection having the $q$-vector of $(\frac{1}{4},-\frac{1}{2},\frac{3}{2})$ from the reciprocal lattice point of $(0,1,-1)$. %
We also observed several magnetic reflections having the same distance from the $c^*$ axis. %
They are magnetic Bragg peaks arising from multiply twinned crystal domains in which the directions of the $c^*$ axes are common to each other, indicating that the sample is not an ideal single crystal. %  
In Fig. \ref{IntMap}(d), we show a line profile along the $(\delta,1-2\delta,\frac{1}{2})$ direction in the 4SL phase \cite{comment1}. %
A sharp peak at the commensurate position of $\delta=\frac{1}{4}$ is consistent with the four-sublattice antiferromagnetic order. %

In the FE-ICM phase, we observed two peaks as indicated by the two blue arrows in Fig. \ref{IntMap}(b). %
From the line profile shown in Fig. \ref{IntMap}(e), the peak positions are determined to be $\delta=0.203$ and $0.297$. %
The former is assigned as a satellite reflection from the reciprocal lattice point of $(0,1,-1)$ using the $q$-vector of $(q_{\rm ICM},-2q_{\rm ICM},\frac{3}{2})$ where $q_{\rm ICM}=0.203$. %
The latter peak can also be assigned as a satellite reflection from $(\frac{1}{2},0,2)$, which becomes a crystallographic zone center due to the lattice distortion associated with the magnetic order\cite{Terada_14.5T,Terada_LatticeModulation}, using the same $q$-vector. %
The two peaks at $\delta=q_{\rm ICM}$ and $\frac{1}{2}-q_{\rm ICM}$ and their intensity distributions are quite similar to those in CuFe$_{1-x}$Ga$_x$O$_2$ with $x=0.035$ in zero magnetic field\cite{Ga-induce}, confirming that the nonmagentic-substitution-induced FE-ICM phase is equivalent to the field-induced FE-ICM phase in pure CuFeO$_2$. %
 
In the $\frac{1}{5}$-plateau phase, a sharp magnetic Bragg reflection was observed at $(\frac{1}{5},\frac{3}{5},\frac{1}{2})$, which corresponds to the commensurate $q$-vector of $(\frac{1}{5},-\frac{2}{5},\frac{3}{2})$, as shown in Figs. \ref{IntMap}(c) and \ref{IntMap}(f). %
This means that there are commensurate three-up-two-down modulations running along the [110] and its equivalent directions on the triangular lattice plane. %
The strong magnetic Bragg peak on the $(h,k,\frac{1}{2})$ plane indicates that the majority of the spins maintain the antiferromagnetic coupling along the $c$ axis. %
This is consistent with the magnetic structure model proposed in the previous study by Mitsuda \textit{et al.}\cite{Mitsuda_2000}, which is shown in the inset of Fig. \ref{IntMap}(f). %

The slopes of the long pulsed magnetic fields can be tuned by changing the charging voltage. %
Figure \ref{Hdep}(a) shows a profile of a long pulsed magnetic field generated by a charging voltage of 142 V. %
Though the maximum of the field is reduced to 9 T, the field variation becomes slower, so that we can measure the diffraction patterns upon the phase transition between the FE-ICM and 4SL phases in detail. %
We repeated the field application with $V_c=142$ V for 40 times, and extracted the diffraction profiles of \#3, \#4, and \#5 frames as shown in Figs. \ref{Hdep}(b), \ref{Hdep}(c), and \ref{Hdep}(d), respectively. %
Among them, the frame \#4 clearly shows the coexistence of the magnetic Bragg peaks of the FE-ICM and 4SL phases as shown in Fig. \ref{Hdep}(c). %
The field variation within this frame is approximately 0.2 T, which is small enough to capture the phase coexistence during the first-order phase transition between the two phases. %

We also performed the measurements with different charging voltages, and obtained the positions of the magnetic peaks on the $(\delta, 1-2\delta,\frac{1}{2})$ line as functions of magnetic field, as shown in  Fig. \ref{Hdep}(e). %
Here, we refer to the in-plane magnetic modulation wavenumber as $q$. %
The $q$-values of the 4SL and $\frac{1}{5}$-plateau phases coincide with the commensurate values, while that of the FE-ICM phase continuously changes with  magnetic field. %
This is consistent with the results obtained by the previous neutron diffraction measurements with steady magnetic field, in which the $q$ values were deduced from the higher-harmonic magnetic reflections on the $(h,k,0)$ plane \cite{Mitsuda_2000,Petrenko_2000}. %
The present results also demonstrate that the stroboscopic neutron diffraction with long pulsed fields is suitable for investigating field-dependent incommensurate magnetic modulations, such as spin-density-wave states of frustrated quantum spin systems in high magnetic fields \cite{Masuda_LiCuVO4}. %

\subsection{CuFe$_{1-x}$Ga$_x$O$_2$ ($x=0.035$)}

The field-induced phase transition in the $x=0.035$ sample was also investigated in the same manner. %
We applied pulsed magnetic fields generated by the charging voltage of 254 V at 4 K, and repeated the field application for 80 times. %
In addition, we also carried out the same set of the measurements with a different delay time to increase the number of data points with different magnetic fields. %
 
\begin{figure*}[t]
\begin{center}
\includegraphics[clip,keepaspectratio,width=18.0cm]{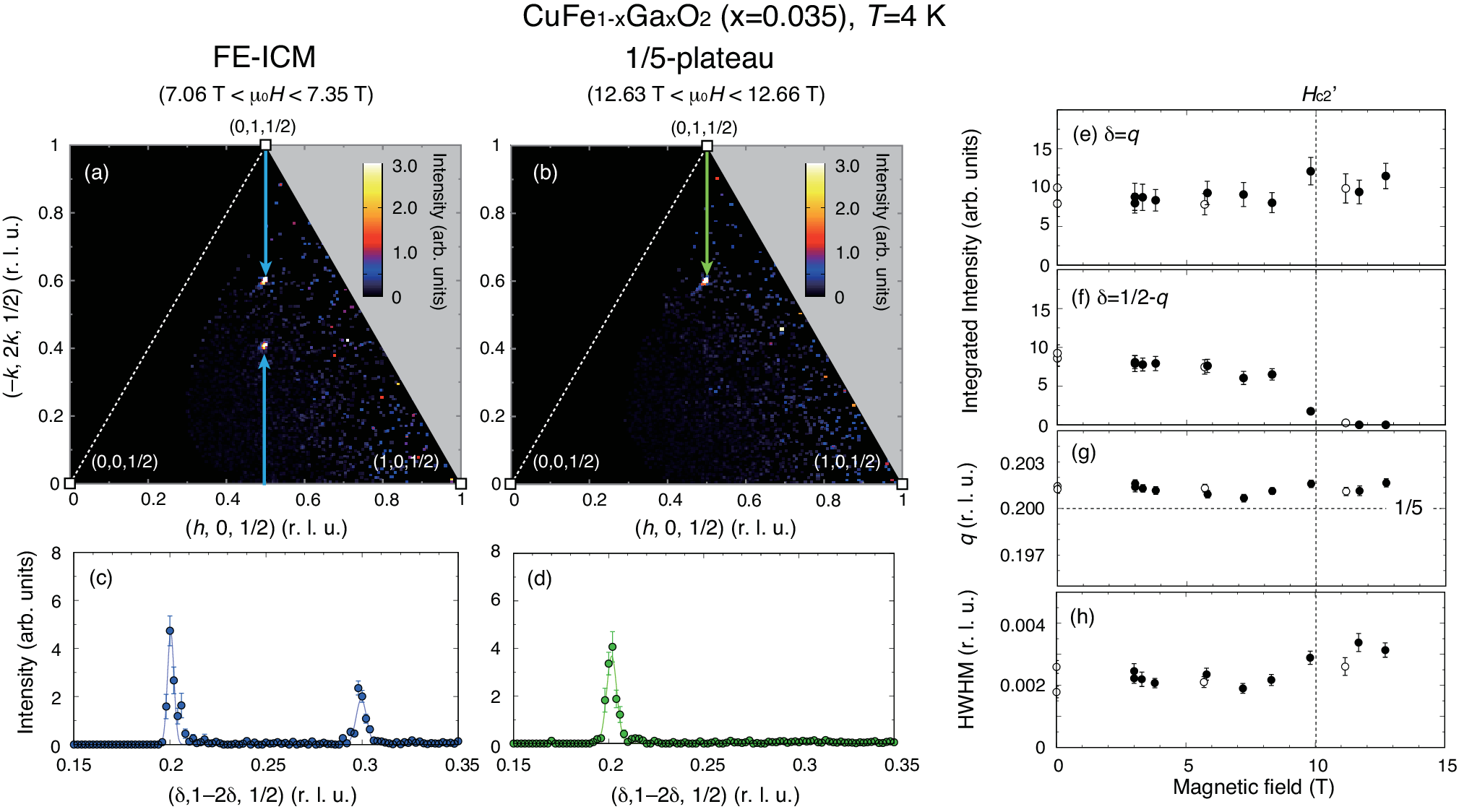}
\caption{Stroboscopic neutron diffraction results on CuFe$_{1-x}$Ga$_x$O$_2$ with $x=0.035$. [(a)-(b)] Intensity maps on the $(h,k,\frac{1}{2})$ plane (a) above and (b) below the critical field $H_{c2}'$. %
These data are obtained by applying the long pulsed magnetic field generated with the charging voltage of 270 V for 80 times at 4 K. %
Solid arrows indicate the positions of the magnetic Bragg peaks in each phase. %
[(c)-(d)] Line profiles along the $(\delta,1-2\delta,\frac{1}{2})$ direction (c) above and (d) below the critical field $H_{c2}'$. %
[(e)-(h)] Magnetic field dependence of the integrated intensities at (e) $\delta=q$ and (f) $\frac{1}{2}-q$, (g) the magnetic modulation wavenumber $q$, and (h) the half width at the half maximum of the peak at $\delta=q$. %
Open and close symbols denote the data measured field-increasing and decreasing process, respectively. }
\label{IntMap_CFGO}
\end{center}
\end{figure*}

Similarly to the results of the $x=0$ sample, we observed a pair of magnetic Bragg reflections on the $(h,k,\frac{1}{2})$ plane in the FE-ICM phase, as shown in Fig. \ref{IntMap_CFGO}(a). %
We note here that the $x=0.035$ sample did not contain any crystal twinning in contrast to the $x=0$ sample. %
Therefore, the signal-to-noise ratio was much improved in both the intensity map and the line profile along the $(\delta,1-2\delta,\frac{1}{2})$ direction shown in Fig. \ref{IntMap_CFGO}(c). 

The critical field of the FE-ICM phase in the $x=0.035$ sample ($H_{c2}'$) is approximately 10 T, above which we found that the peak at $\delta=\frac{1}{2}-q$ disappeared as shown in Figs. \ref{IntMap_CFGO}(b) and \ref{IntMap_CFGO}(d). %
We obtained integrated intensities of the magnetic Bragg peaks at $\delta=q$ and $\frac{1}{2}-q$ as functions of magnetic field as shown in Figs. \ref{IntMap_CFGO}(e) and \ref{IntMap_CFGO}(f), respectively, confirming that the intensity at $\delta=\frac{1}{2}-q$ drops to zero at $H_{c2}'$. %
The diffraction pattern above $H_{c2}'$ is quite similar to that in the $\frac{1}{5}$-plateau phase in pure CuFeO$_2$; in fact, the phase diagram established by previous bulk measurements suggested that the $\frac{1}{5}$-plateau phase appears as the first-field-induced phase in the $x=0.035$ sample\cite{Seki_CFGO,CFGO_highfield_ESR,CFGO_heatcapacity}. %
However, the present results show that the wavenumber $q$ was not exactly on the commensurate position of $\frac{1}{5}$ as shown in Fig. \ref{IntMap_CFGO}(g). %

One possible explanation for the incommensurate nature would be the effects of magnetic domain walls, which may shift the phase of the magnetic modulations. %
A similar peak-shift effect was reported in a commensurate Ising antiferromagnet \cite{CoNb2O6_peakshift}, and became more significant when the system was divided into small magnetic domains. %
Actually, the half width at the half maximum (HWHM) of the peak at $\delta=q$ slightly increases in the field induced phase as shown in Fig. \ref{IntMap_CFGO}(h), which implies the magnetic correlation length becomes shorter upon the field-induced phase transition. %

\section{CONCLUSIONS AND OUTLOOK}

We have established the stroboscopic TOF neutron diffraction measurements with long pulsed magnetic fields at low temperatures. %
The widths of the magnetic fields are more than one order of magnitude larger than the time scale of a polychromatic neutron pulse passing through the sample. %
As a result, we can utilize a wide range of neutron wavelength to measure diffraction patterns in high magnetic fields. %
We applied this technique to study the field-induced magnetic phase transitions of the triangular lattice antiferromagnets CuFe$_{1-x}$Ga$_x$O$_2$ ($x=0, 0.035$), and successfully observed the field evolutions of the magnetic Bragg reflections up to 14 T. %
As for the $x=0$ sample, we directly observed the magnetic Bragg reflections in the field-induced FE-ICM phase, confirming that the peak positions and intensity distribution are the same as those in the nonmagnetic-substitution-induced FE-ICM phase \cite{SpinNoncollinearlity,CompHelicity}. %
We also observed the diffraction patterns of the first field-induced phase of the $x=0.035$ sample, which were found to be quite similar to that in the $\frac{1}{5}$-plateau phase of the $x=0$ sample.  %
The tiny peak shift might be explained by the phase shift of the magnetic modulations at the magnetic domain walls, which is inferred from the broadening of the peak profile. 

One of the biggest advantages of this technique is that one can analyze the neutron diffraction data in the same manner as the analysis for the steady-field measurements. %
Each neutron data frame contains a wide range of wavelength, which enables us to explore the three-dimensional reciprocal space. %
This feature is essential to explore unknown magnetic phases in high magnetic fields. %
It would also be possible to quantitatively investigate field variations of diffuse scattering patterns up to high magnetic fields. %

Another advantage is the potability of the experimental setup. %
As shown in Fig. \ref{cryo}, the cryostat and magnet are relatively small, and can be moved from one beamline to another. %
The capacitor bank and voltage charger are also relatively small; they are approximately 80 cm in width and less than 2 m in height. %
In addition, the stray field form the magnet is also small; specifically, it is less than 3 Oe in a distance of 20 cm from the center of the magnet when the field reaches 14 T. %
Taking into account that the pulsed magnetic field is shielded by the Al-radiation shield and Al-vacuum chamber, it does not affect external instruments. 
These features provide opportunities of high-field experiments for a variety of instruments and their subject matter. %
In fact, high-field environments are necessary not only for physics but also material science and its applications; such as developments of ferromagnetic materials for industrial and medial uses \cite{Oba_pureFe_SANS,KOBAYASHI2023170410}. %

The highest magnetic field of the present setup is 14 T. %
However, it was already demonstrated that long pulsed magnetic fields with the height of 24.3 T and the width at the half maximum of more than 100 ms can be generated by using EDLC banks \cite{LongPulse_EDLC}. %
Therefore, it is possible to develop a long pulsed magnet for neutron scattering whose highest magnetic field exceeds that of existing superconducting magnets for scattering experiments. %
Although a major drawback of the pulsed magnetic field is the relatively long time-interval for repeating the field applications, a recent study shows that the interval can be reduced by using a liquid-helium-cooled coil made of high-purity copper wires, which remarkably reduces Joule heating effect \cite{Long-pulse_He-cool}. %
These technological developments will extend the high-field capability of neutron scattering. %

\section*{Acknowledgements}
The authors thank Y. Ihara for enlightening discussions. %
This work was partly supported by Grants-in-Aid for Scientific Research (Grant No. 22H00104 ) from JSPS. 
The neutron scattering experiment at J-PARC MLF was carried out along the proposals (No. 2021B0070, 2022L0301). %
 
\vspace{5mm}

%\clearpage

\bibliography{LongPulseMag}

\begin{thebibliography}{42}
\bibitem [{com()}]{comment1}%
  \BibitemOpen
  \bibinfo {note} {To obtain the line profiles, the magnetic intensities from
  the misoriented crystal domains are projected on the
  $(\delta,1-2\delta,\frac{1}{2})$ direction taking into account the rotation
  about the $c$ axis.}\BibitemShut {Stop}%
\end{thebibliography}

\end{document}